# Social Trust as a solution to address sparsity-inherent problems of Recommender systems

Georgios Pitsilis
Q2S, NTNU
O.S. Bragstads plass 2E
NO-7491, Trondheim, Norway
+47 735 92743

pitsilis@q2s.ntnu.no

Svein J. Knapskog
Q2S, NTNU
O.S. Bragstads plass 2E
NO-7491, Trondheim, Norway
+47 735 94328

knapskog@q2s.ntnu.no

## ABSTRACT
Trust has been explored by many researchers in the past as a successful solution for assisting recommender systems. Even though the approach of using a web-of-trust scheme for assisting the recommendation production is well adopted, issues like the sparsity problem have not been explored adequately so far with regard to this. In this work we are proposing and testing a scheme that uses the existing ratings of users to calculate the hypothetical trust that might exist between them. The purpose is to demonstrate how some basic social networking when applied to an existing system can help in alleviating problems of traditional recommender system schemes. Interestingly, such schemes are also alleviating the cold start problem from which mainly new users are suffering. In order to show how good the system is in that respect, we measure the performance at various times as the system evolves and we also contrast the solution with existing approaches. Finally, we present the results which justify that such schemes undoubtedly work better than a system that makes no use of trust at all.

## 1. INTRODUCTION
Services offered by recommender systems tend to be hosted in centralized systems. Beside the benefit that is offered in terms of easiness in managing the resources and the availability of the services, there are issues with regard to how much the new users can receive the benefits of their participation in the system. As new users we consider those who have not contributed enough data to the system and hence makes it difficult for predictions for them to be made. Similarly, the same problem seems to exist with items which the users have not have much experience yet. Recommender system services may be offered by social networking platforms like FaceBook [1] and web-pages like dealtime.com [2] or Amazon [3] and may be using mechanisms of User-Based recommender systems for working out predictions for items that users potentially like.

Any potential solution for alleviating the data sparsity issue should not work at the expense of performance of such system but instead it should provide some substantial benefits to users during their bootstrapping. The inability of new users in supplying sufficient quantities of data to the system for predictions to be computed accurately is described in the literature as *"cold-start problem"*[4].

Our approach for overcoming the above problem is based on the idea of extending the neighboring base of new users so that they can be correlated with more participants, not necessarily linked directly with each other via similarity relationships, but been discovered via "friends" as trustworthy for contributing useful data. The trust for them can be inferred via their similar, and hence common, neighbors in a scheme that is known as 'web-of-trust'. In this way, due to the propagation characteristics of trust, it is plausible that similarity between entities that could not be linked previously is becoming exploitable. Users who can be discovered via friends-of-friends might be useful as they may carry valuable experience about some product that is of interest to somebody else. As trust is mainly used for extending the number of relationships between people, users can now cooperate with more participants than before and thus get access to more recommendations. For short we call our system 'hybrid' as it combines both trust networks and traditional recommender systems approaches. We used a framework called "Subjective Logic" for the reasoning of the virtual trust relationships, and since the adaptation to existing recommender system models is a key issue, we used a model that we built our own for inferring trust from the existing users' experiences.

The evaluation we present shows the twofold benefit of this approach as the accomplished reduction of sparsity is accompanied by improved performance. To show the improvement with regard to the *cold-start* problem we demonstrate how some performance metrics evolve as the user community is growing. The rest of the paper is organized as follows: Section 2 is referred to the description of the problem. In section 3 we present the idea and the logical reasoning behind it. An evaluation of the idea follows in section 4, analysis of work that has been done in the area there is in section 5 and finally we conclude with a discussion of the results.

## 2. PROBLEM STATEMENT
The sparsity-inherent problems of recommender systems are related to the fact that a satisfactory number of inferences for either users or items can not be extracted, due to lack of gathered information.

User-Based recommender systems that employ a technique called Collaborative filtering (CF) (in which the user preference for some item is computed upon the similarities between the users), mainly use Resnicks's [10] formula (Equation. 1) for working out predictions $r_{a,}(i)$ for user *a* and items *i*. With $w_{a,u}$ is denoted the Pearson's similarity of user *a* with user *u*, and $r_u$ the rating of user *u* for the item that is of interest to *a*. The formula does not give satisfactory accuracy when sparse datasets are used, as the predictions are highly sensitive to the number of similar participants, *n*.

$$r_a(i) = \bar{r}_a + \sum_{u=1}^{n} w_{a,u} (r_u(i) - \bar{r}_u) \qquad (1)$$

Even though the exploitation of social networks has been recognized as a potential solution for addressing such problems [18], the applied solutions, at least as it seems, do not fulfill this requirement completely. For example, in Epinions.com [17], a well known commercial recommender system, the formation of the trust network is done explicitly by asking users themselves to express whom they trust. In our opinion that is very unproductive for two reasons: first, because not all users are familiar with the notion of trust and hence they are unable to explicitly express whom they trust, and second because people, as it happens for item ratings, are unwilling to invest much time and effort in contributing with their opinions. The latter is considered as the main reason for having sparse datasets.

The *cold-start* problem has been approached in the past from the aspect of being a problem in social networking and recommender systems. In [20] it is suggested that special criteria should be used for deciding to whom it is best for the new users to connect. The decision in this case is based upon users' so far objective assessment of candidates for their suitability and it is decided on how active they have been in contributing data to the system. Nevertheless, such a decision is based solely on quantitative criteria and in our opinion it would be best if qualitative criteria (such as how useful such recommendations have been found to be by the cold start users) were also used. In addition, in the above mentioned solutions the social trust is not adequately exploited, as the discovery for trustworthy participants is usually done by applying local criteria.

It is crucial that when a new idea is applied to an existing system for enhancing its performance it is done in such a way that adapts best to it. Therefore in our model, no additional data should be required to be supplied by users, but the inferred trust should be implicitly derived from the existing evidence instead. This procedure is also useful from the practical aspect, since by doing so the existing recommendation production cycle would not need to change significantly to include the benefits that *social networking* provides. Similar ways of implicit derivation of user's trust from evidence has been proposed for other purposes before, like the EigenTrust algorithm [21] that was mainly build for peer-to-peer systems. However, the trust in that case was essentially perceived as a global reputation value due to being independent on the point of view.

To our knowledge, the use of trust networks for alleviating sparsity-inherent problems, such as the *cold-start* problem in recommender systems have not been adequately studied so far.

With regard to trust models and frameworks, there are many developed so far, most of them mainly to meet special requirements of particular problems, and other more generic ones such as *Subjective Logic* [11]. We mention this framework explicitly because of the substantial adoption it has received for studying the effects of trust propagation in user communities.

## 3. DESCRIPTION OF THE APPROACH

User-based recommender systems that use Resnick's formula are limited to computing predictions of ratings for users whose similarity $w_{a,u}$ with the all contributing participants is known. The limited number of neighbors that can contribute during the system bootstrapping is a significant constraint for achieving good performance during the early stages of the recommender system's life.

Since the accuracy of predictions that the querying user receives is dependent on the number of neighbors/predictors that appear to be similar to him/her, it means that a substantial improvement can be achieved if multiple predictors could be involved in the computation of $r_{a,}(i)$. As new users do not have enough experiences to contribute during the bootstrapping it means the performance would be sub-optimal as there would not be enough links between them.

One characteristic pitfall of a conventional recommendation system mechanism is the inability to incorporate prediction ratings of other participants who have experienced some item that is of interest to the querying agent, but their similarity with the querying agent cannot be inferred.

If a recommender system were to be represented graphically, the similar users would appear to be within a distance of one hop away from the querying user. Exploitation of information that resides at longer distances would be plausible if similarity could have propagative characteristics. Since trust is known for providing such property, which is the key idea of social networks, it means inferring trust from similarity could make it possible to overcome the above limitation. Pursuing this idea further, the neighboring base of users could be extended beyond the one hop range by introducing a hybrid system in which the similarity could be inferred from trust.

The above requirement for predicting the rating that some user *i* would give to item *b* can be expressed as follows:

$$b \in B \mid \{\exists a_j \in n(a_i) : r_i(b) = \perp \land r_j(b) = \perp\} \\ \land \{\exists a_k \in n(a_j) : r_k(b) \neq \perp\} \qquad (2)$$

where *B* is the set of all items in the system, *i,j,k* are 3 users, *a* is the set of users and *n()* is a function that denotes a similar neighbor to *i*.

In a typical scenario of operation of a User-Based recommender system (see figure 1[1]) we assume that Alice and Bob have both experienced a number of items $B_a$ and $B_b$ respectively and hence it can be known how similar they are. On the other hand Clark has another set of common experiences with Bob, and hence it can also be known how similar Bob is with him. However, Clark's experience about some new item that might be of interest to Alice, is not exploitable in the conventional system (Alice's similarity to Clark's is not computable).

---

[1] The letters A…L represent the items rated by users and the numbers 1…5 the rates given.

Extending this consideration and inferring trust implicitly from the calculated similarity for every pair of similar users, then finally a web-of-trust for social partners linked together can be developed.

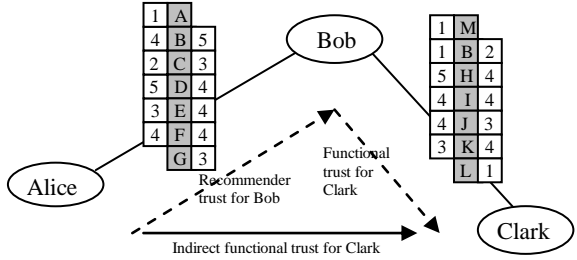

**Figure 1. A typical recommendation production.**

One important issue that comes from the implicit derivation of trust is concerned with the representation of the existing evidence into trust metrics. For that reason some appropriate mapping is necessary.

## 3.1 Trust Modeling

In socially aware systems, users benefit from their trust and connections with others as they can find people they need through the people they trust. Links to a person imply some amount of trust for this person. The importance of social networks is found in the exploitation of such network data to produce information about trust between individuals which have no direct network connection. In theory about trust, this requirement is described with a property called *transitivity* which we are attempting to exploit in this work. As trust is not perfectly transitive in the mathematical sense, however it can be useful in the way like in the real world people consider recommendations from others they trust for their choices. Since trust that derives though recommendations is dependent on the point of view it means it is merely subjective.

In contrast to similarity, trust relationships can be propagated transitively throughout the network of users. In this way the common neighbors can act as trustworthy participants for users of whom similarity is not known, but can be approximated via their derived trust. The concept of computing the indirect trust for distant entities requires the employment of some suitable algebra such as *Subjective Logic*. However, it is required that evidence has first been transformed into some form that the trust algebra can use.

Trust in subjective logic is expressed in a form that is called *opinion* and is referred to a metric that originally was introduced in Uncertain Probabilities theory [22], an extension to probabilistic logic. An *opinion* expresses the belief about the truth of some proposition which may represent the behavior of some agent. The ownership of the opinion is also taken into account and this is what makes the assessment of trustworthiness subjective. Furthermore, this theory is suitable for modeling cases where there is incomplete knowledge. In the case of recommender systems the lack of knowledge about some agent's rating behavior comes from the fact that there are usually limited observations of the rating behavior of some person. The lack of knowledge is actually what shapes the subjective trust or distrust towards that entity. The absence of both trust and distrust in opinions is expressed by the *uncertainty* property. The subjective logic framework uses a simple intuitive representation of uncertain probabilities by using a three dimensional metric that comprises belief (b), disbelief (d) and uncertainty (u) into opinions. It is required that evidence comes in such a form that opinions $\omega_p^A = \{b_p^A, d_p^A, u_p^A\}$ about some agent A with regard to the proposition *p* can be derived from it, and thus be better manageable due to the quite flexible calculus that the opinion space provides. By convention the following rule holds for $b,d,u$: $b+d+u=1$.

*Subjective Logic* provides the following two operators: *recommendation* (3) and *consensus* (4). Both can be used for combining opinions and deriving recommendations regarding other agents in the social network.

$$\omega_p^{A,B} = \omega_p^A \oplus \omega_p^B = \{b_p^{A,B}, d_p^{A,B}, u_p^{A,B}\} \quad (3)$$

A, B are agents and $\omega_p^B = \{b_p^B, d_p^B, u_p^B\}$ is the opinion of B about p expressed as a recommendation to A. $\omega_B^A = \{b_B^A, d_B^A, u_B^A\}$ is the opinion of A about the recommendations of B. The consensus opinion $\omega_p^{AB}$ is held by an imaginary agent AB representing both A and B.

$$\omega_p^{AB} = \omega_B^A \otimes \omega_p^B = \{b_p^{AB}, d_p^{AB}, u_p^{AB}\} \quad (4)$$

The output values $b,d,u$ of the combined opinions are derived from simple algebraic operations. More about this can be found in [11]. In our opinion the above considerations of Subjective Logic are quite sufficient for deriving recommendations with regard to the rating behavior of users whose subjective trustworthiness can be computed transitively within the social network of trusted participants.

In the literature trust is also distinguished into *direct* and *indirect trust*, the former when it is derived from personal experience of a trustor, and the latter when it is derived from recommendations of others. Also, another distinction of it is *functional trust*, which expresses the trustworthiness for some agent with regard to some proposition p, and *recommendation trust* which expresses the trustworthiness of some agent as a recommender.

In our example depicted in figure 1, Bob has *functional trust* in Clarke's rating behavior, but the trust that can be derived by Alice for Clark via Bob's recommendation trust is *indirect trust* since Alice does not have her own evidence to support it, but merely trusts Bob's taste. Finally, the trust that Alice is interested in knowing for Clark, is actually an *indirect functional trust,* which indicates how much she would trust him for his taste.

As far as the evidence transformation is concerned, various models for converting ordinary observations into evidence have been proposed [11][19]. For our approach, we have used a simple model which is best suited to recommender system data [12]. In our work we have come up with a solution of deriving the trustworthiness that a pair of agents would place in each other by using existing data such as ratings for items they have gathered experience with. In the same work an approach for mapping trust into similarity is also introduced. This is explained below.

For calculating the uncertainty we used the simplified formula: $u = (n+1)^{-1}$, in which $n$ denotes the number of common experiences in a trust relationship between two agents A and B.

The derivation of opinions from existing user experiences with items can be done by using an appropriate formula such as the one given below which we used for our experiment. This formula was used for shaping the belief property (*b*) of *Subjective Logic* from *User Similarity* ($W_{a,u}$) also known as *Correlation Coefficient*.

$$b = \frac{1}{2}(1-u)\left(1 + W_{a,u}^{\kappa}\right) \quad (5)$$

The disbelief property *d* of the opinions can easily be derived from the remainder of *b* and *u* as: *d =1-b-u*.

In our case scenario of recommender system the calculated belief *(b)* is referred to either *recommender* or *functional* trust. In equation 5 the k value denotes the exponent in the equation used for transforming the similarity metric into derived belief (k=1 for linear transformation).

In the figure below we present pictorially a high level view of our hybrid system that could take advantage of this idea. The trust derivation mechanism for predicting the ratings that users would give to products can be easily embedded into the existing ordinary system.

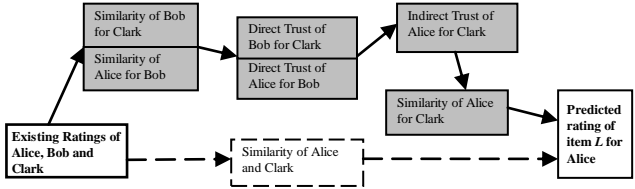

**Figure 2. The conceptual view of our hybrid system**

The part of the diagram shown in dotted line represents the existing recommendation production mechanism and it applies only if evidence suffices for computing the similarity of Alice to Clark.

## 4. EVALUATION

We evaluated the performance of our hybrid approach against various alternatives such as the standard CF technique which employs the Pearson similarity and uses Resnick's prediction formula [10].

The results which demonstrate the effectivenes of predicting user likeness express the ability of the system to identify potentially unsatisfactory options. Moreover, we introduce a set of metrics to demonstrate efficiency in terms of sparsity reduction as well as effectiveness against the *cold-start* problem.

For the evaluation we used a subset of a movie recommendation system called Movielens. This original dataset contains more than 1 million movie recommendations submitted during a period of 812 days by around 6000 users for 3100 movies. To capture the dynamic characteristics of performance that evolve over time, we made use of the timestamp (TS) information that is attached on every submitted rating. We divided the rating experiences into 5 sets, each containing ratings submitted within the same period of time. Finally, we performed the experiment for each of those sets separately. As the number of recommendations performed at each stage is more important to be shown than the timestamp information we considered as the best solution to present the adjacent *sparsity* value. That is the percentage of ratings in the users by items matrix for which originally no values had been provided.

To fulfill the requirement for calculations of similarity to be stable we assumed that similarity between two users is calculable only if there are at least 10 items that have been rated by both of them.

It is worth mentioning that in the current experiment, no control has been applied on filtering the number of neighbors that a user can have, and the only limiting factor for forming a trust relationship is the number of common experiences that exist.

### 4.1 Metrics Used

*Predictive Accuracy* metrics such as MAE are quite popular for measuring how close the recommender system's predicted ratings are to the true user ratings [13]. However, it is also interesting to know how good the system would be in successfully identifying the items that users would be unhappy with, and therefore we also used *Classification Accuracy* metrics. To justify this decision we claim that in the way the experiment was done, there were no rating predictions attempted for items that had no recorded rating experiences and hence the danger that one might be lead into classification errors is significantly reduced.

#### 4.1.1 Measuring Coverage

With *Coverage* we refer to the percentage of items for which predictions can be made and in [14] it is defined as:

$$C = \frac{\sum_{a_i \in A}\left|\{b \in B \mid \exists a_j \in n(a_i) : r_j(b) \neq \perp\}\right|}{|B| \cdot |A|} \quad (6)$$

where A and B are the set of users and products respectively, $r_j$ is a rating function and $n(a)$ denotes the set of similar neighbors of some user *a*.

We introduce one new metric, *User Coverage Gain* (UCG), to demonstrate the actual benefit that users receive when they make use of the trust graph. This metric relates the cost $|A|$ with the benefit R, expressed as a ratio of the hybrid system by the standard CF. It can be computed using the following formula:

$$UCG = \left[\frac{R_h}{|A_h|} \cdot \left(\frac{R_s}{|A_s|}\right)^{-1} - 1\right]_{TS} \quad (7)$$

$R_h$ and $R_s$ refer to the number of predictions that the system was capable of performing for the new users in timestamp TS for the hybrid and the standard recommender system respectively. $|A_h|$ and $|A_s|$ refer to the sizes of populations of users which have made use of the trust network for discovering other participants at time *TS,* and correspondingly the adjacent number of users who would use the standard CF for performing those predictions. The populations of users are expressed as in formulas (8) and (9). The formula for $A_h$ is corresponding to the scenario that trust propagation has been restricted to max distance of 2 hops only, which is the case for our experiment. The items recommended in every timestamp can be expressed as in formula (10)

$$A_s = \sum_{b \in B} |\{a \in A | \exists c \in n(a) : r_a(b) = \bot \land r_c(b) \neq \bot \}| \quad (8)$$

$$A_h = \sum_{b \in B} |\{a \in A | \exists c_k \in n(a), \exists d \notin n(a) \land d \in n(c_k) : r_a(b) = \bot \land r_d(b) = \bot \}| \quad (9)$$

$$\sum_{a_i \in A} |\{b \in B : r_i(b) \neq \bot \}| \quad (10)$$

For showing the level of contribution of the trusted participants to a prediction to be made, we came up with a metric called *Trust Graph Contribution*. This metric is presented in formula (11) and as can be seen it relates the relative increase in the number of users (due to the use of trust network) used for the produced recommendations, with the actual number of recommendations produced. This relative increase is expressed as the ratio of *trusted neighbors* by all neighbors (trusted and similar). With *"trusted neighbors"* we refer to those users in the social graph for which their similarity with the querying user was derived by propagated trust recommendations and was not calculated directly by correlating the common experiences (ratings) with some neighbor as it is done for the case of "similar" ones.

$$Ctrb = \left( |R|^{-1} \cdot \sum_{i \in R} \left[ \frac{|A_h|}{|A_h| + |A_s|} \right] \right)_{TS} \quad (11)$$

In formula (11), $A_s$ and $A_h$ are the same as explained before, $R$ is the set of recommendations produced in some timestamp and is the sum of $R_h$ and $R_s$. This metric demonstrates how effective the discovery of neighbors of interest can be via the social network and it is interesting to know how it contributes in the performance improvement.

Beside new users, items for which there extensive experience is not available are also affected by the *cold-start* problem. Therefore, we found it necessary to measure the improvements that the hybrid solution would offer to them as well.

### 4.1.2 Measuring Accuracy

*Predictive Accuracy* is a standard metric for measuring how close the predicted ratings are to the true user ratings. In our experiment we measure the MAE (Mean Average Error) which shows the absolute deviation between the two. However, as MAE can be unimportant for showing the performance for items of interest to users, we considered also using other accuracy metrics in addition to this.

*Classification Accuracy* metrics are used to measure the frequency with which the system makes incorrect or correct decisions about whether an item is bad or good and it is usually applied in connection with the task of finding lists of top items.

As we mentioned previously we consider it more appropriate to demonstrate how useful the system is in helping users to avoid making choices of products that they might be unhappy with. Therefore we used the metric *F-score*, also called *Harmonic Mean*, and it is used in *information retrieval*. F-score measures the effectiveness of retrieval with respect to the cost of retrieving the information [13].

It is necessary that the negativ*e* (N) and positive (P) instances are clearly distinguished, and for our particular case we characterize as N the case of experiences with products that the user would be unsatisfied with and would give low rating. The opposite case corresponds to P. *Precision* is defined as: $P = \frac{TP}{TP + FP}$ and represents the ratio of instances that were correctly predicted as non-satisfactory by the user against all instances that were predicted as non-satisfactory. *Recall* is: $R = \frac{TP}{TP + FN}$ and represents the number of instances that were correctly predicted as non-satisfactory ones normalized by the total number of instances that actually received unsatisfactory rating. The *F-score that* shows the relative tradeoff between the benefits (TP) and the costs (FP) is calculated using the formula: $F = \frac{2PR}{P + R}$.

As rating values of 1 and 2 represent an unfortunate choice and 4 and 5 a successful choice, we used the value 3 as threshold for considering an experience as unsatisfactory. In table 1 we present the confusion matrix. We considered as true positives (TP) the instances that where correctly classified as receiving low rating and false negatives (FN) those instances that were classified as having high rating, but still predicted as been non-satisfactory to the users. True negatives (TN) denote those which were correctly classified as giving a high rating. Finally, false positives (FP) are referred to the number of bad items which were mistakenly classified by the system as satisfactory ones for the new users.

**Table 1. Confusion Matrix**

| Actual \ Predicted | Predicted Value ≤ 2 | Predicted Value > 2 |
|---|---|---|
| Rating ≤ 2 | TP | FP |
| Rating > 2 | FN | TN |

The evaluation was done using the cross validation technique *leave-one-out* applied on every user rating. The process was the following: every rating provided by each user was removed from the dataset and then its value was tentatively computed using the trust network. The computed and the removed value are then compared and the error was calculated. The evaluation algorithm is presented in the figure below.

| **Algorithm:** Evaluation plan |
|---|
| 1. **for** all users *i* who have provided at least 10 ratings |
| 2. **for** all items *k* of user *i* |
| 3. Pset ← ø |
| 4. **for** all users j whose common rated items with *i* > 10 |
| 5. **if** ( *j* similar to *i* ) |
| 6. Pset ← Pset ∪ { *j* } |
| 7. *Similar* ← *Similar* +1 |
| 8. **else if** ( trust of *i* for *j* is computable) |
| 9. **derive** similarity of *i* for *j* from trust of *i* for *j* |
| 10. Pset ← Pset ∪{ *j* } |
| 11. *Trusted* ← *Trusted* +1 |
| 12. **end for** |
| 13. **predict** *k* for over Pset |
| 14. **calculate** MAE for *k* |
| 15. **calculate** *TP,FP,NF,F-score for i* |
| 16. **end for** |
| 17. **end for** |
| 18. *Trust Graph Contribution* ← *Trusted* / (*Trusted* + *Similar* ) |
| 19. **average** MAE for all *i* |

**Figure 3. The evaluation plan in pseudo-code**

With regard to coverage, only those values which were possible to compute were considered for contributing to it.

To study more closely the benefits that our system can offer to the entities that are mostly affected by the cold-start problem, we repeated the experiments considering the new items and the new users alone. To achieve that for every timestamp we filtered and counted those entities which committed their first experience at that particular timestamp.

## 4.2 Results – Discussion

### 4.2.1 Overall Performance

With regard to the overall performance of the system, we first demonstrate the results we obtained for the *Coverage Gain* and the *Contribution of Trust Graph*. In these diagrams, time is represented by its adjacent sparsity value and is shown across the horizontal axis.

In figure 4 is shown how the *Contribution of the Trust Graph* develops over the experiment. For every recommendation produced, the ratio of trusted participants considered for this recommendation divided by all participants (trusted and similar) used for the same recommendation is counted and the results are averaged. As can be seen from the diagram this metric follows in general a decreasing trend throughout the simulation, but more importantly, it gets its maximum value towards the beginning when the trust network is still building up. We interpret this as a good indication that our system can cope well with the cold start problem as the resources of the hybrid system are shown to be exploited better at that timestamp. The decreasing trend followed afterwards is explained as the effect of gradual replacement of trust relationships by computed similarity as more and more recommendations are submitted over time. In the same figure is shown the benefit of using the hybrid system in terms of *User Coverage Gain* in comparison to using the standard Collaborative Filtering. Interestingly enough, this metric follows an increasing trend and more importantly it remains unaffected by the decreasing rate of new users.

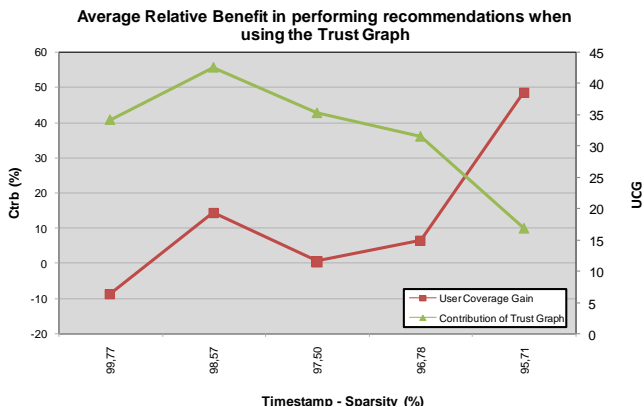

**Fig. 4. The Coverage Gain & Trust graph Contribution**

That is because the new users who join late get higher support as the friends-of-friends network is then denser than before. As far as the rating accuracy is concerned the results for both the classification accuracy (F-score) and predictive accuracy (MAE) are shown in table 2 and are also graphically presented in figure 5.

For the classification accuracy, the results show that there is quite a notable advantage of our method over the standard CF in discovering those items that a user would be unhappy to choose.

**Table 2. Accuracy expressed in F-score and MAE**

| Timestamp (sparsity) \ Method | F-score | | MAE % | |
|---|---|---|---|---|
| | Standard | Hybrid | Standard | Hybrid |
| 1- (99,77 %) | 0,0836 | 0,0952 | 15.944 | 15.686 |
| 2- (98,57 %) | 0,1546 | 0,1832 | 14.562 | 15.126 |
| 3- (97,50 %) | 0,2443 | 0,2720 | 14.652 | 15.350 |
| 4- (97,78 %) | 0,3030 | 0,3350 | 15.228 | 15.676 |
| 5- (95,71 %) | 0,3347 | 0,3598 | 15.210 | 16.100 |

As can be seen this advantage appears from early on (first timestamp), it maximizes at the second timestamp (highest difference between the F-score values of the standard and hybrid models) when the data is still quite sparse, and continues so at all consecutive timestamps. It is interesting to note that this behavior is very unlikely to be coincidental as it appeared at all five different datasets we tested.

One can also see that the predictive accuracy appears to be higher in the proposed system than in the standard one, but here there is temporary decrease during the early timestamps.

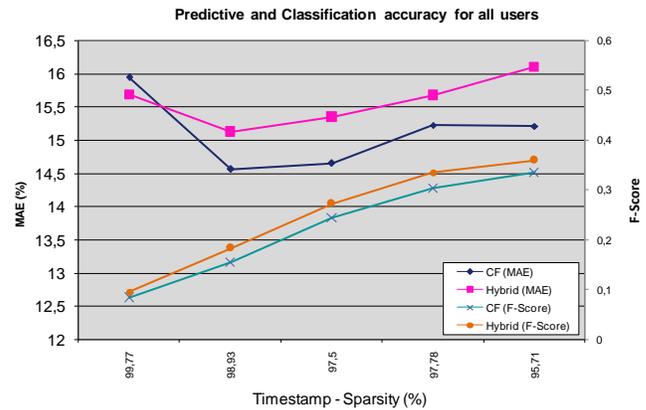

**Figure 5. The accuracy for the compared schemes**

### 4.2.2 Selective Performance

Next we present our results which demonstrate the behavior of the examined systems when considering only the new users and items. First we demonstrate the sizes of populations of *cold-start* users and items that appeared for the first time on each timestamp.

From the diagrams in figure 6 it can be seen that in the case of the hybrid system, the cold-start users are shown to be committing their first experience with the hybrid system earlier than in the standard CF. In the first two timestamps, the number of new users in the former is higher than in the latter. As expected though, that trend is declining as the system develops over time. This looks quite reasonable from the way our experiment was done, as in contrast to a real world running system, we used restricted size sets of 100 users.

The early emergence of new users that appears in the hybrid system is indicative of its success in exploiting the social network and performing predictions that wouldn't be possible in the conventional one, and hence attract more users. Consequently the same happens for *cold-start* items which are now discovered and rated by users earlier in the hybrid system than in the standard recommnder system.

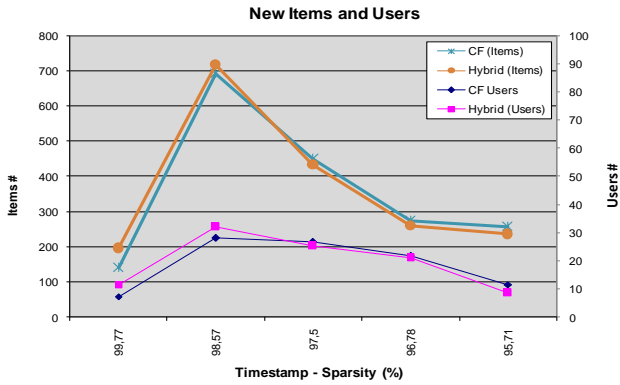

**Figure. 6. The new users and items**

In table 3 we present the prediction and classification accuracy for *cold-start* users and items and the results are pictorially presented in figures 7 and 8. As far as prediction accuracy is concerned, from the results it can be seen that the deployment of the trust system into the existing one has no impact on the accuracy of ratings prediction, as the error is kept low (below 15%) during the early stages of the system. For the new items the situation looks quite a lot better as there is no noticeable penalty in the prediction error against using the standard recommender system. In comparison to the overall performance results of figure 5, the new users receive higher benefit (MAE 14.87%) than the average user (MAE 15.13%) during the early timestamps (at TS:2). However, this benefit is diluted as the time progresses. Instead, new users loose this advantage if using the standard system. (MAE:14.56% opposed to 14.58%).

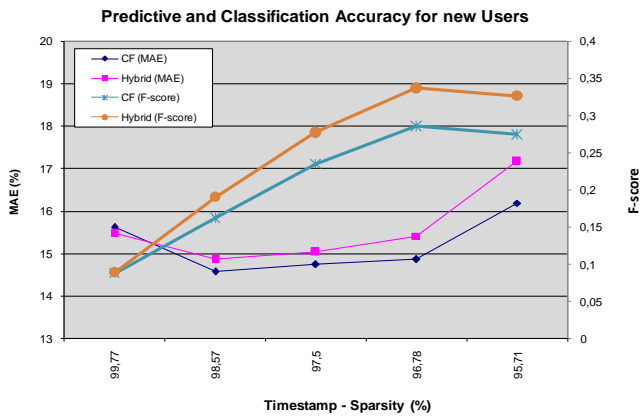

**Figure. 7. The benefit of hybrid on new users**

Finally, it is also important to note the increasing trend in the average error as seen in fig. 7 which means that new users who join the system late are less likely to receive good service than those who join early.

Regarding classification accuracy for new users and new items, our measurements show that the proposed hybrid system outperforms the traditional one at all time instances.

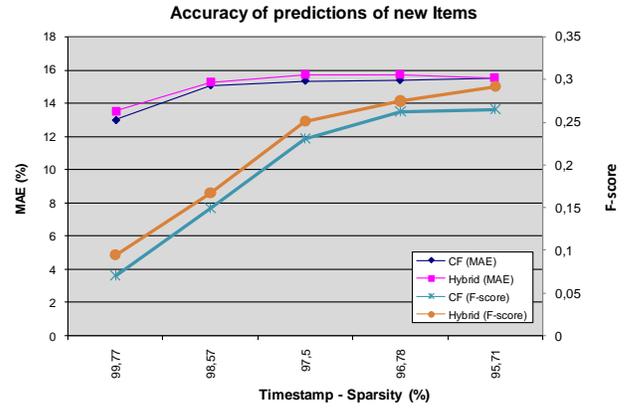

**Figure. 8. The benefit of hybrid system on new items**

The growing trend for F-score as it appears in fig. 7 is indicative of the increasing benefit that new users receive as the system develops. In comparison to the classification performance for all users presented in fig. 5, the new users receive higher benefit than anyone else as they are potentially guided better to avoid products they will be unhappy with.

We consider the above two observations as a positive consequence for the proposed system for compensating the users early on. It is important that new users receive the highest benefit as they are assumed to be less tolerant in receiving poor recommendations. Experiencing poor recommendations consistently over time may reduce their trust towards the system and make them reluctant to rely on it for delivering good service. If the original trust disappears, the users interest in using the system may vanish altogether.

**Table 3. Accuracy for new Users and new Items.**

| Predictive accuracy in MAE | | | | |
|---|---|---|---|---|
| **Method Timestamp (sparsity)** | New Users | | New Items | |
| | Standard | Hybrid | Standard | Hybrid |
| **1 - (99,77 %)** | 15.62 | 15.8 | 13.00 | 13.52 |
| **2 - (98,57 %)** | 14.58 | 14.87 | 15.06 | 15.27 |
| **3 - (97,50 %)** | 14.75 | 15.04 | 15.32 | 15.71 |
| **4 - (97,78 %)** | 14.86 | 15.42 | 15.37 | 15.72 |
| **5 - (95,71 %)** | 16.19 | 17.18 | 15.52 | 15.51 |

| Classification accuracy in F-score | | | | |
|---|---|---|---|---|
| **1 - (99,77 %)** | 0.088 | 0.090 | 0.071 | 0.095 |
| **2 - (98,57 %)** | 0.162 | 0.190 | 0.149 | 0.167 |
| **3 - (97,50 %)** | 0.235 | 0.277 | 0.230 | 0.250 |
| **4 - (97,78 %)** | 0.286 | 0.337 | 0.262 | 0.274 |
| **5 - (95,71 %)** | 0.275 | 0.326 | 0.265 | 0.291 |

## 5. BACKGROUND RESEARCH

Trust has been the subject of investigation by many researchers in the past for alleviating issues connected with the use of sparse datasets in recommender systems. Singular Value Decomposition has been proposed by other researchers and found to be better than the standard collaborating filtering [5] for alleviating sparsity problems. Other approaches are based on the idea of removing global effects and estimating the interpolation weights for each weighting factor for improving the accuracy of recommender systems [6]. Hybrid systems which combine content and collaboration have also been proposed in which various weights are set on the contribution of similarity [7]. In such an approach, the weight is dependent on the number of common items. In [15], O'Donovan and Smyth study the effects of using trust models in the recommendation process and they demonstrate how it behaves against various attack scenarios. In [8], a solution for computing trust in CF systems has been investigated, but in the proposed model the trustworthiness of the recommender is not taken into account. In [9], in the work done by Lathia et.al., it is suggested that collaboration groups could better be formed by k-trusted neighbors rather than k-similar ones. In [16], the cold-start problem is approached using some idea based on machine learning. Massa et al. in [23] has published a similar idea with ours, but based on different working hypothesis which requires that users would provide the trust statements themselves. To our knowledge trust has not been studied adequately so far as a solution to the cold-start problem. In addition, even though all the studies performed can demonstrate the advantages of using trust, they are merely static and do not capture the characteristics of the community as it evolves. Since the cold-start problem is a time related issue we chose to demonstrate our proposed solution in a way that it can be shown if the advantage actually becomes available when the system needs it the most.

## 6. CONCLUSION

We have proposed a hypothetical hybrid recommender system which uses trust to exploit the latent relationships between users and we have measured its performance. In this way, also knowledge that exists at distant participants can be discovered and used by users who do not need to be known to each other. We used our modeling technique to build trust from existing evidence. The evaluation results show a significant benefit against the standard technique both in terms of coverage and in accuracy of predictions. It is interesting to note that the benefit is more distinguishable for new users and items which traditionally are mostly affected by the sparsity problem. Furthermore, the higher values achieved for F-score are indicative of improved ability in protecting users from choosing products that they may not like. With regard to the challenge of alleviating the cold-start problem, it can be seen that the benefits of using the trust enabled system are particularly visible early on when they are actually needed. A future challenge is to extend even further the period of time that the benefit is received.

## REFERENCES

[1] Facebook Social Network service, http://www.facebook.com

[2] General Consumer Review Site, http://www.dealtime.com

[3] Electronic Commerce Company, http://www.amazon.com

[4] Maltz D., Ehrlish K., Pointing the Way: Active Collaborative filtering, In proc. of CHI-95,pp.202-209,New York, ACM Press (1995)

[5] Sun X., Kong F., Ye S., A Comparison of Several Algorithms for Collaborative Filtering in Startup Stage, In proc Networking Sensing and Control, IEEE, pp.25-28 (2005)

[6] Bell R., Koren Y., Improved Neighborhood-based Collaborative Filtering, In proc IEEE International Conference on Data Mining, pp.7–14 (2007)

[7] Melville P., Mooney R.L. , Nagarajan R., Content-Boosted Collaborative Filtering for Improved Recommendations, In proc of Eighteenth national conf. of Artificial Intelligence, pp.187-192, ISBN:0-262-51129-0 (2002)

[8] Quercia D., Heiles S., Carpa L., B-trust: Bayesian Trust Framework for Pervasive Computing. In proc 4th International Conf. iTrust 2006. Lecture Notes in Computer Science (Vol.3986/2006). Springer, pp. 298-312 (2006)

[9] Lathia N., Hailes S., Carpa L., Trust-Based Collaborative Filtering, in proc IFIPTM, Springer, Vol. 263, pp.119-134, Trondheim, Norway (2008)

[10] Resnick P., Varian H.R., Recommender Systems, Communications of the ACM. 40(3), pp. 56-58 (1997)

[11] Jøsang A., A Logic for Uncertain probabilities, International Journal of Uncertainty, fuzzi-ness & Knowledge based systems,Vol.9, No.3 (2001)

[12] Pitsilis G., Marshall L.F., Modeling Trust for Recommender Systems Using Similarity Metrics, in proc. IFIPTM, Springer, Vol. 263, pp.103-118, Trondheim, Norway (2008)

[13] Herlocker, J. L., Konstan, J. A., Terveen, L. G., and Riedl, J. T. 2004. Evaluating collaborative filtering recommender systems. *ACM Trans. Inf. Syst.* 22, 1 (Jan. 2004)

[14] Ziegler C-N., Towards Decentralized Recommender Systems, ISBN 363901149X, Vdm-Verlag, (2008)

[15] O'Donovan J., Smyth B. ,Is trust Robust?: An Analysis of Trust-Based Recommendation, in Proc. of 11th International Conf. on Intelligent User Interfaces IUI '06, pp.101-108, (2006).

[16] Xuan N. L., Thuc V., Trong D. L., Anh D. D., Addressing cold-start problem in recommendation systems, in Proc. of the 2nd international conf. on Ubiquitous information management and communication, Jan.31-Feb.01, Suwon, Korea, (2008).

[17] General consumer review site. http://www.epinions.com,

[18] Golbeck J., Hendler J.. Inferring Trust Relationships in Web-Based Social Networks, *ACM Transactions on Internet Technology,* 6(4). (2006)

[19] Josang A.,Bhuiyan T.,Xu Y.,Cox C., Combining Trust and Reputation Management for Web-Based Services, in Proc. of the 5th international conf. on Trust, Privacy and Security in Digital Business,Turin, Italy, pp.90-99 (2008)

[20] Victor P., Cornelis C., Teredesai A. M., De Cock M., Whom should I trust?: the impact of key figures on cold start recommendations , In proc. SAC '08: ACM symposium on Applied computing, pp. 2014-2018. (2008),

[21] Kamwar S. D, Schlosser M. T, Hector Garcia-Molina. "The EigenTrust algorithm for reputation management in P2P networks". In: Proceedings of the 12th International Conference on World Wide Web, Budapest, Hungary, 640-651 (2003)

[22] Shafer G., A Mathematical Theory of Evidence, Princeton University Press (1976)

[23] Massa P., Avesani P.,Trust Aware Bootstrapping of Recommender Systems, in proc. ECAI Workshop on Recommender Systems (2006).